\documentclass[a4paper,11pt]{article}
\usepackage{jinstpub} 
\usepackage{lineno}
\usepackage{siunitx}
\DeclareSIUnit{\niel}{\mbox{\SI{1}{\mega\electronvolt}\ n$_{\mathrm{eq}}$\ \si{\per\cm\squared}}}
\newcommand{\mnka}{Mn-K$_{\alpha}$ }
\newcommand{\mnkb}{Mn-K$_{\beta}$ }
\newcommand{\sika}{Si-K$_{\alpha}$ }
\newcommand{\sikb}{Si-K$_{\beta}$ }
\newcommand{\sikab}{Si-K$_{\alpha, \beta}$} 
\newcommand{\siesc}{Mn-K$_{\alpha, \beta}$ - Si-K$_{edge}$ }

\title{\boldmath Yield and performance validation of the Monolithic Stitched Sensor (MOSS), the first wafer-scale prototype for the ALICE ITS3 upgrade}

\collaboration[c]{on behalf of the ALICE collaboration}

\author[a,b]{M. W. Menzel}
\affiliation[a]{European Organisation for Nuclear Research (CERN), Meyrin, Switzerland}
\affiliation[b]{Ruprecht Karl University of Heidelberg, Heidelberg, Germany}

\emailAdd{marius.wilm.menzel@cern.ch}

\abstract{
The ALICE Inner Tracking System upgrade (ITS3) will employ stitched, wafer-scale Monolithic Active Pixel Sensors (MAPS) for the first time in high-energy physics, achieving a material budget of only \SI{0.09}{\percent X_{0}} per layer. Its first stitched prototype, the Monolithic Stitched Sensor (MOSS), underwent serial testing confirming sensor yield compliance with ITS3 requirements. In-beam tests show the device meets the ITS3 efficiency requirement of >\SI{99}{\percent} while maintaining a fake-hit rate below \SI{0.1}{hits/pixel/s}, with performance sustained up to irradiation levels of \SI{4}{\kilo\gray} and \SI{4e12}{\niel}. The sensor demonstrates excellent charge-collection properties and linearity between time-over-threshold and deposited energy in the \SIrange{1.8}{6.5}{\kilo e \volt} range in response to soft X-ray emissions. This article provides an overview of the validation steps and characterisation results.}


\notoc
\begin{document}\maketitle
\flushbottom
\section{Introduction}

Monolithic Active Pixel Sensors (MAPS) are employed in high-energy physics detectors to provide excellent spatial resolution while maintaining a low material budget. The ALICE experiment exploited these advantages with the ALPIDE sensor in the Inner Tracking System 2 (ITS2). Its upgrade, ITS3 \cite{its3}, aims to reduce the material budget of the three innermost half layers from \SI{0.36}{\percent X_{0}} to \SI{0.09}{\percent X_{0}} per layer, with the remaining contribution arising mostly from the sensor itself. To achieve this, the ITS3 design foresees six wafer-scale MAPS, each up to \qtyproduct{265 x 98}{\mm} in size, thinned down to \SI{50}{\micro\meter} thickness, and bent into cylindrical half layers with radii of \SI{19.0}{\milli\meter}, \SI{25.2}{\milli\meter}, and \SI{31.5}{\milli\meter}, respectively. They are supported by low-density carbon foam, and have a target power consumption of \SI{40}{\milli\watt\per\cm\squared} allowing for air cooling. For ITS3, a detection efficiency above \SI{99}{\percent} with a fake-hit rate below \SI{0.1}{hits/pixel/s} is required. This performance must be maintained up to an ionising radiation dose of \SI{4}{\kilo\gray} and a non-ionising dose of \SI{4e12}{\niel}. To assess the feasibility and performance of such devices in view of the ITS3 design, the MOnolithic Stitched Sensor (MOSS) was fabricated. This article presents the MOSS layout, demonstrates the validation of the pixel-matrix performance, and summarises the yield-assessment results.

\section{The Monolithic Stitched Sensor}

\begin{figure}[htbp]
\centering
\includegraphics[width=.75\textwidth]{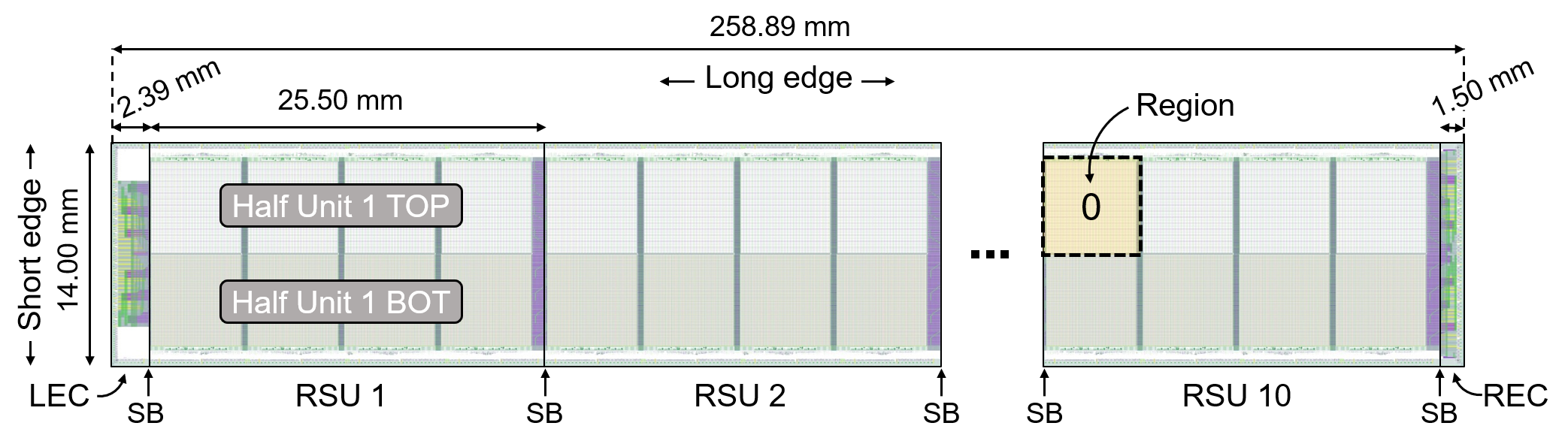}
\caption{Sensor layout, illustrating the modular architecture consisting of Left End-Cap (LEC), Right End-Cap (REC) and Repeated Sensor Units (RSUs) with Stitched Boundaries (SB) indicated. \cite{moss}\label{fig:moss}}
\end{figure}

With the MOSS prototype \cite{moss}, stitching is explored as a method for fabricating wafer-scale MAPS for ITS3. The approach relies on a modular reticle design. In MOSS, the reticle consists of three layout components: the Left End-Cap (LEC), the Right End-Cap (REC), and the Repeated Sensor Unit (RSU). Placed side by side on the wafer, as illustrated in Fig.~\ref{fig:moss}, the components are designed to interconnect via data and power lines at the Stitching Boundries (SB), forming a single MAPS. Each RSU is divided into a top and bottom half unit, each containing four regions. Each region hosts a pixel matrix along with the corresponding biasing, control, and readout units. The top regions host $256\times256$ pixels with a pitch of \SI{22.5}{\micro\meter}, while the bottom regions feature $300\times300$ pixels with \SI{18}{\micro\meter} pitch. The pixel design of the sensor is derived from the earlier DPTS prototype~\cite{dpts}, featuring a deep, low-doped n-well implant for each pixel. Adjacent implants are separated by a gap to enhance the electric field towards the pixel's collection diode, improving charge collection near the pixel boundaries and mitigating charge sharing.

\section{Pixel-matrix performance}

\begin{figure}[htbp]
\centering
\includegraphics[width=.83\textwidth]{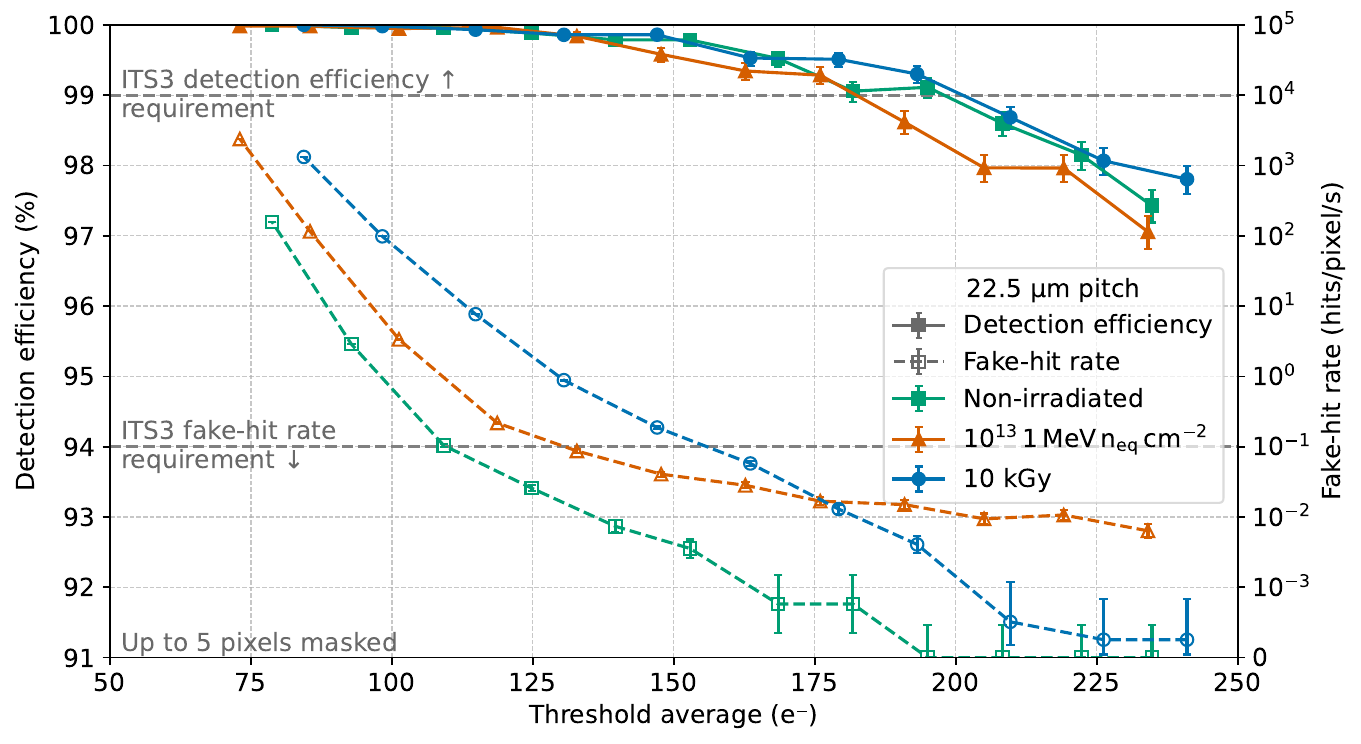}
\caption{Detection efficiency and fake-hit rate as a function of the threshold for a non-irradiated sensor, a sensor irradiated to \SI{10}{\kilo\gray} ionising dose, and a sensor irradiated to \SI{1e13}{\niel} non-ionising dose. All three sensors show a threshold range above \SI{50}{e^{-}} complying with the ITS3 requirements.\label{fig:eff} \cite{moss}}
\end{figure}

The in-beam performance of the pixel matrix was characterised using a reference tracking telescope in a charged-particle beam \cite{moss}. Figure~\ref{fig:eff} presents the detection efficiency and fake-hit rate as functions of the threshold for different irradiation levels. The threshold range in which the ITS3 efficiency requirement (>~\SI{99}{\percent}) and the fake-hit rate requirement (<~\SI{0.1}{hits/pixel/s}) are simultaneously fulfilled is referred to as the operational margin. For a non-irradiated sensor, an operational margin of approximately \SI{90}{e^{-}} is observed, while for irradiated sensors, it reduces to less than \SI{50}{e^{-}}. For the sensor irradiated to a dose of \SI{10}{\kilo\gray} of ionising radiation, an increase in the fake-hit rate is observed. This behaviour is consistent with radiation-induced degradation of the transistor performance, leading to a reduced signal-to-noise ratio in the frontend. For the sensor irradiated to a fluence of \SI{1e13}{\niel}, the observed increase in fake-hit rate can be attributed to enhanced shot noise resulting from elevated leakage current. Furthermore, the fake-hit rate saturates at higher thresholds, consistent with residual radioactivity originating from activation of the sensor material itself. Despite these effects, the sensor meets the efficiency and fake-hit rate requirements even after irradiation to the dose expected for ITS3.

\begin{figure}[htbp]
\centering
\includegraphics[width=.83\textwidth]{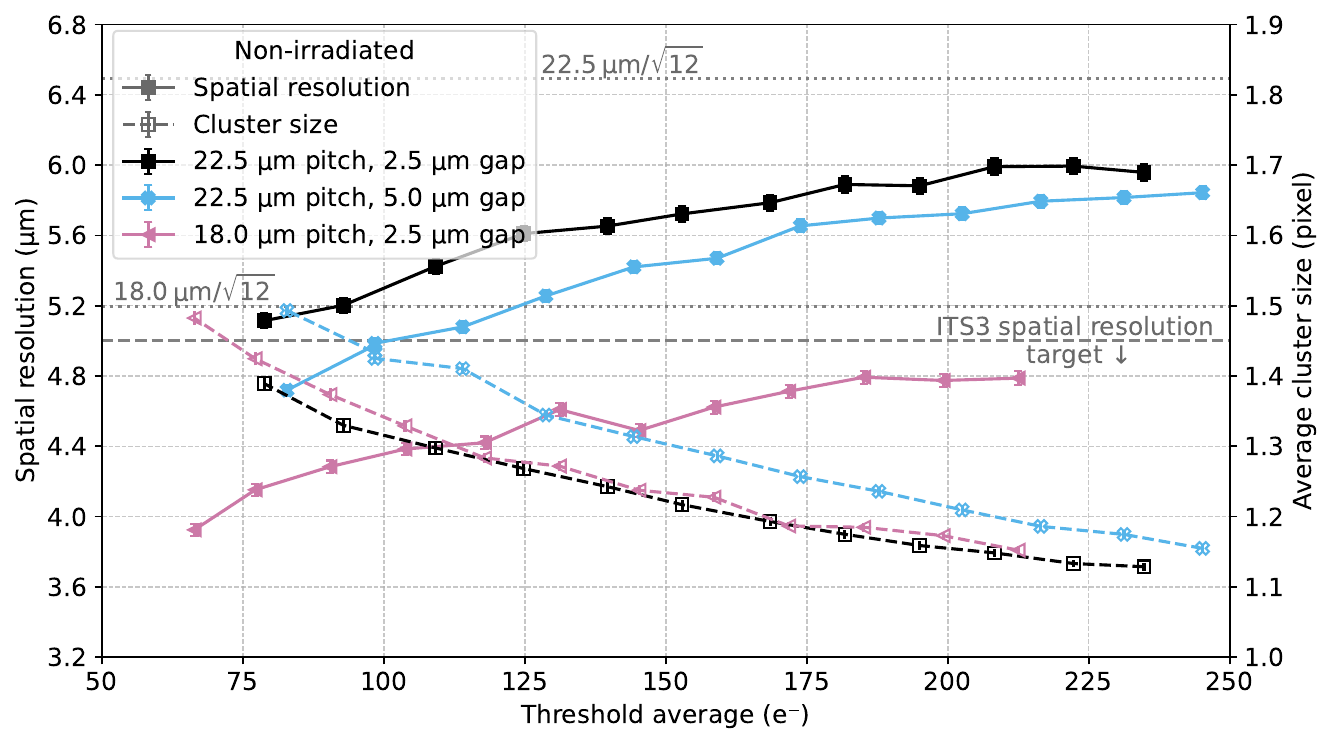}
\caption{Spatial resolution and cluster size versus threshold for three pixel variants, showing the effect of pitch and gap size. At \SI{160}{e^{-}}, the \SI{18}{\micro\meter} pitch variant reaches about \SI{4.5}{\micro\meter} resolution, compared to \SI{5.7}{\micro\meter} for the \SI{22.5}{\micro\meter} pitch. Doubling the gap size improves the resolution by roughly \SI{0.4}{\micro\meter}. \cite{moss}\label{fig:res}}
\end{figure}

When a particle traverses the active area, multiple adjacent pixels can register signals above threshold, forming a cluster. The sensor’s spatial resolution can be determined from the RMS of the distance between the particle track’s intercept on the sensor and the cluster center of mass. From this value, the tracking resolution of the telescope at the position of the tested device (about \SI{2}{\micro\meter}) is quadratically subtracted. The spatial resolution is computed separately for the column and row directions and averaged. Figure~\ref{fig:res} shows the spatial resolution and average cluster size as a function of threshold for different pixel geometries: two variants with \SI{22.5}{\micro\meter} pitch and varying gap sizes (\SI{2.5}{\micro\meter} and \SI{5.0}{\micro\meter}), and a variant with \SI{18}{\micro\meter} pitch. The ITS3 spatial-resolution target of \SI{5}{\micro\meter}~\cite{its3} is indicated, along with the expected binary resolution for each pitch, i.e., the resolution if all charge were always collected by a single pixel. The \SI{18}{\micro\meter}-pitch variant achieves a spatial resolution below the ITS3 target across the full threshold range. At the lowest threshold within the post-irradiation operational margin (\SI{160}{e^{-}}, cf.~Fig.~\ref{fig:eff}), it reaches about \SI{4.5}{\micro\meter}, compared to \SI{5.7}{\micro\meter} for the \SI{22.5}{\micro\meter}-pitch design. Between gap sizes of \SI{2.5}{\micro\meter} and \SI{5.0}{\micro\meter}, a difference of approximately \SI{0.4}{\micro\meter} is observed at \SI{160}{e^{-}}, attributable to increased charge sharing in the larger-gap variant.

\section{Yield estimation of the stitched design}

\begin{figure}[htbp]
\centering
\includegraphics[width=.65\textwidth]{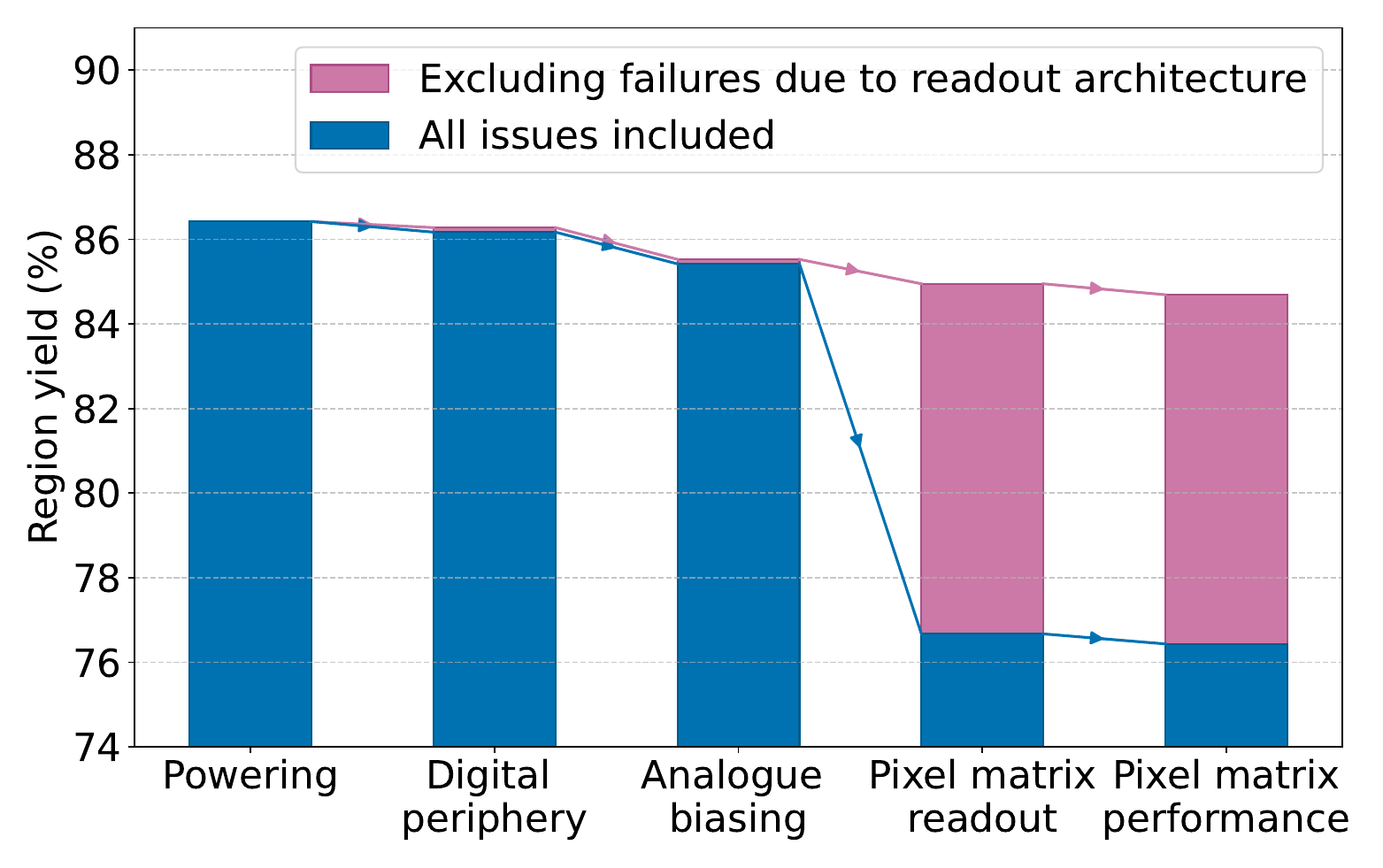}
\caption{Yield losses per failure category, normalised per region. Excluding readout-architecture failures, a yield of about \SI{85}{\percent} is achieved. When additionally excluding powering issues not relevant to the final sensor design, the effective region yield exceeds \SI{98}{\percent}.\cite{moss}\label{fig:yield}}
\end{figure}

To assess the production yield of the stitching technology, 82 sensors from 14 different wafers were tested \cite{moss}. The procedure included powering tests, verification of the digital periphery (e.g.\ slow-control functionality and register access), and evaluation of the analogue periphery (e.g.\ linearity of the digital-to-analogue converters). Furthermore, the pixel-matrix readout was tested to verify the correct propagation of a firing pixel through the digital periphery to the acquisition system. Finally, the pixel-matrix performance was evaluated, with regions containing more than \SI{1}{\percent} faulty pixels classified as unsatisfactory.
Figure~\ref{fig:yield} summarises the yield per test category, normalised per region. Approximately \SI{86}{\percent} of regions exhibit satisfactory performance. During powering tests, about \SI{14}{\percent} of regions showed layout-specific shorts in the on-chip power network. A detailed investigation presented in \cite{metal_stack} traced these failures to the novel, customised metal stack of the sensor. Since the metal stack will be replaced in the final ITS3 sensor, these limitations are not expected to impact the ITS3 yield. Additionally, around \SI{7}{\percent} of regions exhibited failures in the pixel-matrix readout. Given that the simplified readout architecture of the prototype was not optimised for fault tolerance and will be replaced by a more robust design in the final sensor, these issues are likewise not expected to affect the final yield. Other failure categories show a robust design with failures below \SI{1}{\percent}. In summary, when excluding known issues not applicable to the final sensor implementation, a region yield above \SI{98}{\percent} is achieved.

\section{Response to soft X-rays}

\begin{figure}[htbp]
\centering
\includegraphics[width=.55\textwidth]{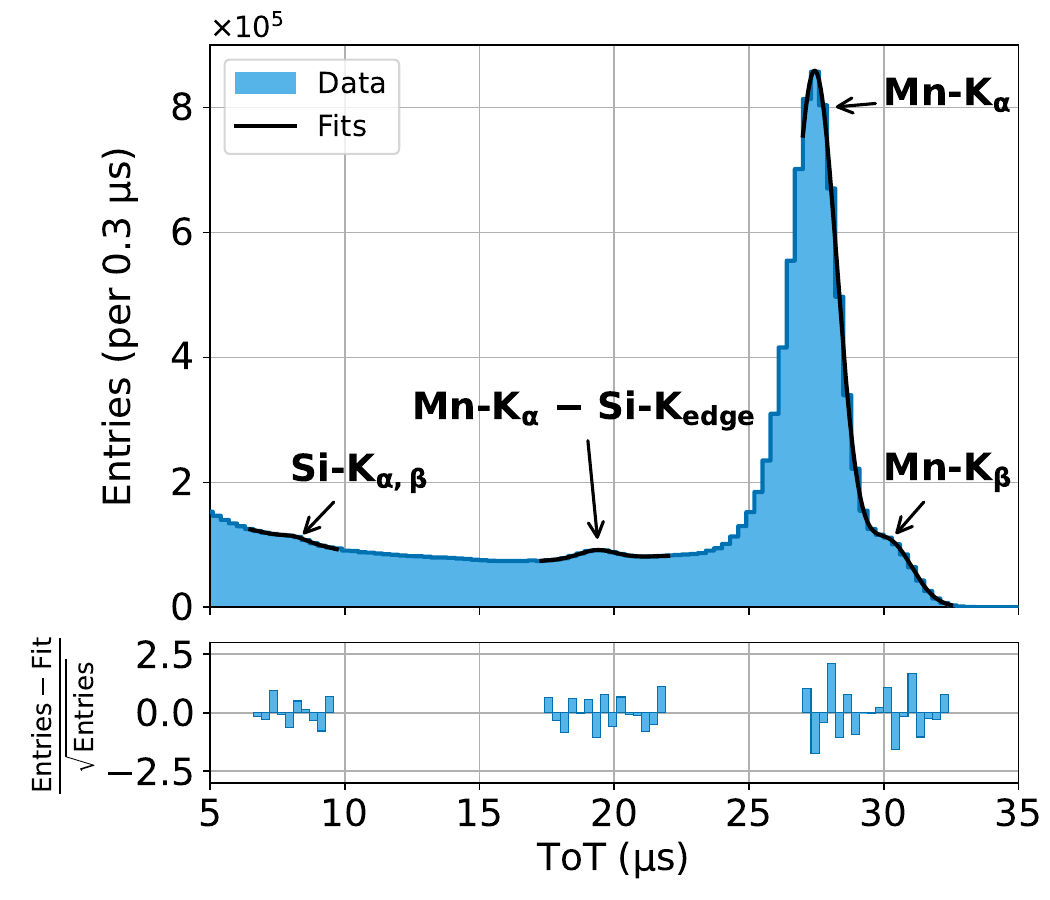}
\caption{Single-pixel cluster ToT spectrum in response to $^{55}$Fe X-ray emissions. The primary \mnka and \mnkb peaks are resolved, along with the secondary \sika and \siesc features originating from X-ray interactions within the silicon. Fit residuals remain within 2.5 standard deviations.~\cite{moss}\label{fig:fe55}}
\end{figure}

\begin{figure}[htbp]
\centering
\includegraphics[width=.55\textwidth]{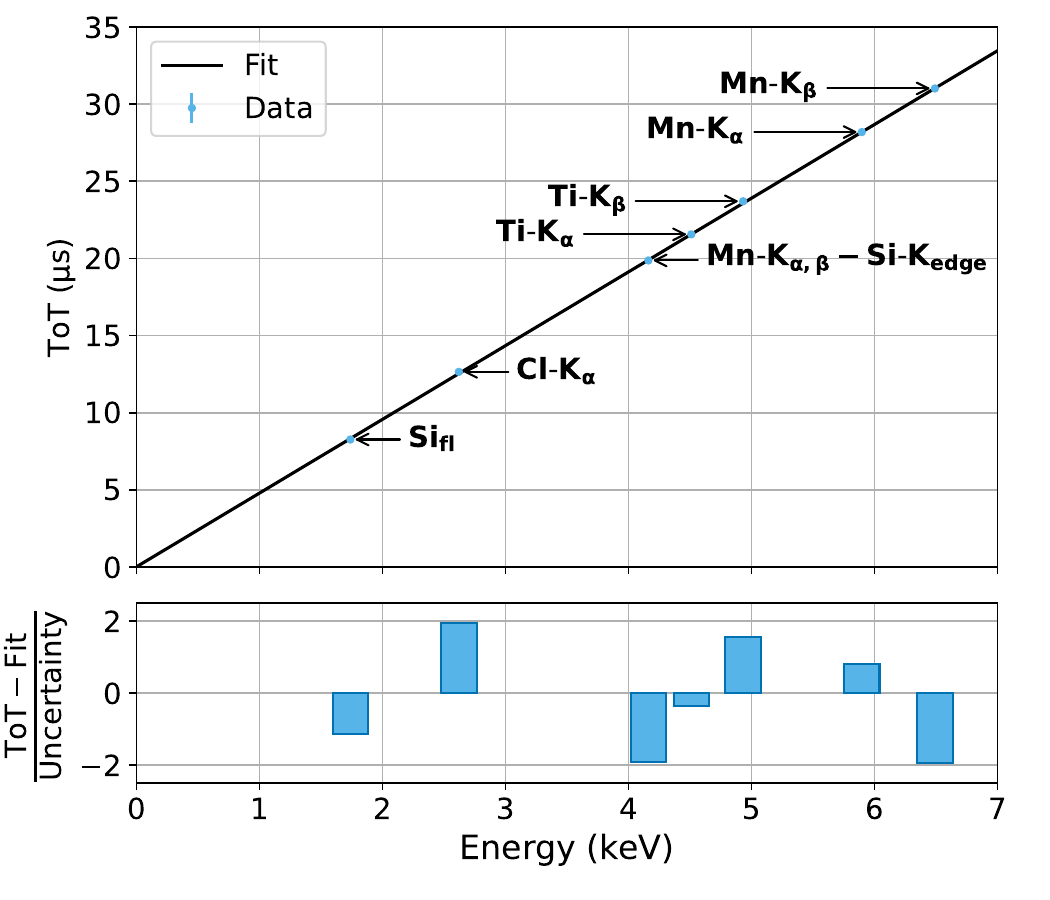}
\caption{ToT response as function of the emission energy. Fit residuals are within two standard deviations, validating the sensor's energy linearity.\label{fig:ecal}}
\end{figure}

The in-pixel amplifier of MOSS is designed such that the Time-over-Threshold (ToT) of a signal is proportional to the energy deposited in the sensor \cite{moss}. Figure~\ref{fig:fe55} shows a ToT spectrum in response to $^{55}$Fe X-ray emissions. With an energy resolution of $\textrm{FWHM}/\textrm{Mean}=\SI{7.3(0.2)}{\percent}$, the \mnka and \mnkb emission lines are resolved. Furthermore, spectral features arising from secondary X-ray interactions within the silicon itself are resolved. First, a photoelectron with an energy of \siesc is emitted from the K-shell of a silicon atom when an incident \mnka or \mnkb X-ray with an energy of \SI{5.9}{\kilo\electronvolt} or \SI{6.49}{\kilo\electronvolt} is absorbed. Second, if the subsequent de-excitation of the atom occurs via an L- or M-shell transition, a \sika or \sikb fluorescence photon is emitted. While the \mnka and \mnkb emissions are fitted by a sum of two Gaussians, the \siesc and \sikab{} emissions are fitted each with a Gaussian added to a linear background. Figure~\ref{fig:ecal} relates the observed peak positions in ToT space with the corresponding energies. The fit validates the linearity of the ToT response in the \SIrange{1.8}{6.5}{\kilo e \volt} range.

\section{Conclusions}

The successful testing of the first stitched MAPS prototype for ITS3 marked a major milestone in the R\&D effort. MOSS was shown to be fully functional and demonstrated a \SI{98}{\percent} regional yield (excluding well-understood issues not relevant to the final device), confirming that the stitching technique can achieve yields compatible with the ITS3 design requirements. The in-beam performance of the pixel matrix was validated in terms of detection efficiency and fake-hit rate up to the ITS3 radiation-hardness specifications. The \SI{22.5}{\micro\meter}-pitch variant achieved a spatial resolution of about \SI{5.7}{\micro\meter} at the lowest threshold satisfying efficiency and fake-hit rate requirements after irradiation (\SI{160}{e^{-}}). For the \SI{18}{\micro\meter}-pitch design, a resolution of about \SI{4.5}{\micro\meter} was obtained, indicating that an intermediate pixel pitch is sufficient to meet the ITS3 spatial-resolution target.
In summary, the MOSS sensor demonstrates the feasibility of employing wafer-scale sensors with sufficient yield and in-beam performance for the ITS3 upgrade. The remaining critical tasks for the next chip include validating high-speed data transmission and managing voltage drops arising from powering the sensors solely via the LEC and REC. The extensive characterisation campaign with MOSS establishes the foundation for the deployment of large-scale stitched MAPS in ITS3 and future detector systems.

\bibliographystyle{JHEP}
\bibliography{biblio.bib}

\end{document}